\documentstyle[epsfig]{mn}
\begin{document}
\LARGE
\normalsize

\title
[]
{The radio luminosity of persistent X-ray binaries}
\author[R. P. Fender and M. A. Hendry]
{R. P. Fender$^{1}$\thanks{email : rpf@astro.uva.nl}
and M. A. Hendry$^{2}\thanks{email : martin@astro.gla.ac.uk}$\\
$^{1}$ Astronomical Institute `Anton Pannekoek' and Center for High
Energy Astrophysics, University of Amsterdam, Kruislaan 403, \\
1098 SJ Amsterdam, The Netherlands\\
$^{2}$ Department of Physics and Astronomy, University of Glasgow,
Glasgow G12 8QQ, Scotland, UK}

\maketitle

\begin{abstract}

We summarise all the reported detections of, and upper limits to, the
radio emission from persistent (i.e. non-transient) X-ray binaries. A
striking result is a common mean observed radio luminosity from the
black hole candidates (BHCs) in the Low/Hard X-ray state and the
neutron-star Z sources on the horizontal X-ray branch.  This implies a
common mean intrinsic radio luminosity to within a factor of twenty
five (or less, if there is significant Doppler boosting of the radio
emission).  Unless coincidental, these results imply a physical
mechanism for jet formation which requires neither a black hole event
horizon or a neutron star surface.  As a whole the populations of
Atoll and X-ray pulsar systems are less luminous by factors of $\ga 5$
and $\ga 10$ at radio wavelengths than the BHCs and Z sources (while
some Atoll sources have been detected, no high-field X-ray pulsar has
ever been reliably detected as a radio source). We suggest that all of
the persistent BHCs and the Z sources generate, at least sporadically,
an outflow with physical dimensions $\geq 10^{12}$ cm,
i.e. significantly larger than the binary separations of most of the
systems.  We compare the physical conditions of accretion in each of
the types of persistent X-ray binary and conclude that a relatively
low ($\leq 10^{10}$ G) magnetic field associated with the accreting
object, and a high ($\geq 0.1$ Eddington) accretion rate and/or
dramatic physical change in the accretion flow, are required for
formation of a radio-emitting outflow or jet.

\end{abstract}

\begin{keywords}

binaries : close --- ISM : jets and outflows --- radio continuum : stars

\end{keywords}

\section{Introduction}
\label{sec:intro}

Radio synchrotron emission is observed from $\sim 20$\% of X-ray
binaries (e.g. Hjellming \& Han 1995). In several cases the radio
emission has been resolved into jet-like outflows reminiscent of the
jet/lobe structures in AGN (e.g. Fender, Bell~Burnell \& Waltman 1997;
Mirabel \& Rodriguez 1999, Fender 2000).  Much recent work, both
theoretical and observational, e.g. Hjellming \& Johnston (1988),
Penninx (1989), Hjellming \& Han (1995), Falcke \& Biermann (1996) and
Livio (1997), has discussed not only these clearly-resolved
relativistic outflows, but also weaker unresolved radio emission from
X-ray binaries.  It seems increasingly possible that all radio
emission from X-ray binaries could arise in such outflows.

However, little serious study has been made of the properties of radio
emission from the persistent sources to see whether it is consistent
with this wide-ranging model. In this paper we consider the radio
luminosity of the persistently accreting BHC, Z-type and Atoll-type
X-ray binaries and discuss whether observations are compatible with
the generic jet picture.

\section{The sample : types of X-ray binary}
\label{sec:types}

In this work we are interested in the radio emission from {\em
persistently} accreting X-ray binaries, i.e. X-ray binary systems in
which we infer from their continual and reliable detection by X-ray
satellite missions that they are in a quasi-steady state of stable
accretion.  Van Paradijs (1995; hereafter vP95), lists 193 X-ray
binaries in the most up to date catalogue available. More than 70 of
these systems are transients with unstable accretion and are not
considered here.  In addition, more than 15 new X-ray binaries have
been discovered since the publication of the catalogue, but again they
are all transient, and not under study in this work.

We shall ignore the distinction between low- and high-mass X-ray
binaries, which is of concern for the evolution of the systems but
probably not for the physics of jet formation and its coupling to the
inner accretion disc.

This leaves four clear subclasses of X-ray binaries : 

\begin{itemize}

\item{Black Hole candidates (BHCs)} 

\item{Z-type (neutron star) X-ray binaries }

\item{Atoll-type (neutron star) X-ray binaries}

\item{X-ray pulsar (neutron star) X-ray binaries}

\end{itemize}

In addition, the majority of systems in the vP95 catalogue are
unclassified beyond being indicated as Bursters and/or Dippers.
However, advances in our knowledge and understanding of the properties
of the neutron-star X-ray binaries indicate that the majority of the
unclassified systems are likely to be Atoll-like, although a small
subclass of lower-luminosity sources is possible (Ford, van der Klis
and van Paradijs, private communication). So, the majority of the
persistently detected X-ray binaries are likely to be Atoll-type
neutron star low-mass binaries.  X-ray transients are generally
low-mass BHCs, although a small number are Atoll-like. The sample of
persistent BHCs (four plus two in the LMC) and Z-type sources (six),
both intrinsically very luminous, is likely to be more or less
complete for our Galaxy and the Magellanic Clouds. We note that Smale
\& Kuulkers (1999) have recently claimed that LMC X-2 may be a Z
source but do not consider it as such in this work and their
interpretation has yet to be confirmed.

Below we shall discuss the state of our knowledge about the radio
emission in these four classes of system.

\begin{table}
\centering
\begin{tabular}{cccc}
\hline
Name & Mean cm radio & distance & Refs\\
     & flux density (mJy) & (kpc)    & \\
\hline
Cyg X-1 & $12.9 \pm 0.4$ & $2.5 \pm 0.5$ & 1,2,3\\
GX 339-4 & $5.0 \pm 2.2$ & $3.0 \pm 1.5$ & 4,5,6,7\\
1E1740.7-2942& $\sim 0.5$ & $8.5 \pm 1.5$ & 8,9,10,11\\
GRS 1758-258 & $\sim 0.5$ & $8.5 \pm 1.5$ & 8,9,10\\
LMC X-1      & $< 1.5$    & $55 \pm 5$ & 12\\
LMC X-3      & $< 0.12$   & $55 \pm 5$ & 12\\
\hline
\end{tabular}
\caption{Observed mean radio flux and 
estimated distances 
for the persistent BHCs 
Refs 
1: Pooley, Fender \& Brocksopp (1998),
2: Gies \& Bolton (1986),
3: Han (1993)
4: Sood et al. (1997),
5: Fender et al. (1997a),
6: Callanan et al. (1992),
7: Hannikainen et al. (1998),
8: Mirabel (1994),
9: Mart\'\i{}  (priv. comm.),
10: Mart\'\i{} (1993),
11: Anantharamaiah et al. (1993),
12: Fender, Southwell \& Tzioumis (1998)}
\end{table}

\begin{table}
\centering
\begin{tabular}{cccc}
\hline
Name & Mean cm radio & distance & Refs\\
     & flux density (mJy) & (kpc)    & \\
\hline
Sco X-1 & $10 \pm 3$ & $2.0 \pm 1.0$ & 1,2 \\
GX 17+2 & $1.0 \pm 0.3$ & $7.5 \pm 2.3$ & 1,3\\
GX 349+2& $0.6 \pm 0.3$ & $5.0 \pm 1.5$ & 4,5\\
Cyg X-2 & $0.6 \pm 0.2$ & $8.0 \pm 2.4$ & 1,6,7\\
GX 5-1  & $1.3 \pm 0.3$ & $9.2 \pm 2.7$ & 1\\
GX 340+0& $0.6 \pm 0.3$ & $11.0 \pm 3.3$ & 8\\
\hline
\end{tabular}
\caption{Observed mean radio flux densities and 
estimated distances and
for the Z-sources. 
Refs 
1: Penninx (1989),
2: Crampton et al. (1976),
3: Penninx et al. (1988),
4: Cooke \& Ponman (1991),
5: Christian \& Swank (1997),
6: Hjellming et al. (1990),
7: Cowley, Crampton \& Hutchings (1979),
8: Penninx et al. (1993).}
\end{table}

\begin{table}
\centering
\begin{tabular}{cccc}
\hline
Name & Mean cm radio  & distance & Refs\\
     & flux density (mJy) & (kpc)    & \\
\hline
GX 9+1 & $< 0.2$ & $8.5 \pm 1.5$ & 1,2,3\\
GX 3+1 & $< 0.3$ & $8.5 \pm 1.5$ & 1,2,3\\
GX 13+1 & $1.5 \pm 0.5$ & $4.5 \pm 1$ & 1,2,3,4\\
GX 9+9 & $<0.2$ & $8.5 \pm 1.5$ & 2,3\\
4U 1820-30 & $<0.2$ & $6.4 \pm 0.6$ & 1,2,3,4,5\\
4U 1705-44 & $<0.2$ & $11.0 \pm 3.3$ & 3,4\\
4U 1636-53 & $<0.2$ & $5.5 \pm 1.7$ & 1,3\\
4U 1735-44 & $<0.2$ & $9.2 \pm 2.8$ & 1,2,3\\
GX 354+0 & $<0.3$ -- $\sim 0.5$ & $5.7 \pm 1.7$ & 6\\
4U 1608-52 & $<0.2$ & $4.4 \pm 1.3$ & 1,4 \\
4U 1837+04 & $<0.4$ & $8.4 \pm 2.5$ & 2,4\\
4U 1702-42 & $<0.4$  & $7.3 \pm 2.2$ & 2,4\\
4U 1850-08 & $<0.3$ & $6.5 \pm 0.6$ & 2,4\\
\hline
\end{tabular}
\caption{Atoll sources
Refs
1 : van Paradijs \& White (1995),
2 : Grindlay \& Seaquist,
3 : Berendsen et al. (1999),
4 : Christian \& Swank (1997),
5 : Mckie (1997),
6 : Mart\'\i{} et al. (1998)
}
\end{table}

\begin{table}
\centering
\begin{tabular}{cccc}
\hline
Name & Mean cm radio   & distance & Refs\\
     & flux density (mJy) & (kpc)    & \\
\hline
GX 1+4 & $< 0.2$ & $6 \pm 1$ & 1 \\
X Per & $< 2.2$ & $0.7 \pm 0.3$ & 2 \\
Her X-1 & $< 1.3$ & $6.6 \pm 1$ & 2,3\\
SMC X-1 & $< 0.2$ & $55 \pm 5$ & 4 \\
LMC X-4 & $< 0.2$ & $45 \pm 5$ & 4 \\
Vela X-1 & $< 0.2$ & $1.9 \pm 0.1$ & 4,5 \\
1E 1048.1-5937 & $< 0.2$ & ? & 4 \\
Cen X-3 & $< 0.2$ & $8 \pm 1$ & 4 \\
1E 1145-614 & $< 0.2$ & $8 \pm 1$ & 4, 6\\
GX 301-2 & $< 0.2$ & $5 \pm 1$  & 4, 5 \\
3A 1239-599 & $< 0.2$ & ? & 4 \\
4U 1626-67 & $< 0.2$ & $8 \pm 1$ & 4, 7 \\
OAO 1657-41 & $< 0.2$ & $\leq 10$ & 4 \\
\hline
\end{tabular}
\caption{X-ray pulsars
Refs 
1. Fender et al. 1997,
2. Nelson \& Spencer 1988,
3. Reynolds et al. (1997),
4. Mckie (1997),
5. Kaper et al. (1997),
6. Ilovaisky et al. (1982),
7. Chakrabarty (1998)
}
\end{table}

\subsection{Black hole candidates}
\label{subsec:BHCs}

Four binary systems in our Galaxy (Cyg X-1, GX 339-4, 1E1740.7-2942
and GRS 1758-258) and two in the LMC (LMC X-1 and LMC X-3) are
considered to contain persistently accreting stellar mass black holes.
The majority of other BHCs are X-ray transient systems which spend
most of the time in the `off' state in which the accretion rate is
very low and the associated X-ray emission very weak.  Two of the
persistent sources (Cyg X-1 and GX 339-4) also undergo state changes,
typically being observed in the `low/hard' state but occasionally
switching to the `high/soft' state, sometimes via `intermediate'
states (e.g. Zhang et al. 1997; Mendez \& van der Klis 1997) but the
dynamic range in observed luminosity is less than that of the
transient systems and they can (almost) always be detected by X-ray
satellite missions.  An `off' state, corresponding to very low X-ray
flux levels, has also been observed from GX 339-4.

The radio emission from Cyg X-1 and GX 339-4 in the low/hard X-ray
state is relatively well studied, being observed to be weak (in
comparison to more extreme systems like Cyg X-3) and steady, and
roughly correlated with the X-ray emission (Pooley, Fender \&
Brocksopp 1999; Brocksopp et al. 1999; Hannikainen et al. 1998; Corbel
et al. 2000).  In addition, the radio emission is observed to drop
below detectable levels during transitions to intermediate or
high/soft X-ray states (e.g. Tananbaum et al. 1972; Fender et
al. 1999a). The radio spectra of these two systems are approximately
flat between 2 -- 15 GHz (Pooley et al. 1999; Fender et al. 1997,
Corbel et al. 1997), with no observed cut-off at high or low
frequencies. Recent mm-wavelength observations have shown that this
flat spectrum continues to at least 220 GHz in Cyg X-1 (Fender et
al. 1999b).  The first imaging of a compact jet from Cyg X-1 has
recently been reported (Stirling, Spencer \& Garrett 1998). Fewer
observations have been made of the Galactic Centre systems
1E1740.7-2942 and GRS 1758-258, but they too appear to be weak and
steady, with approximately flat radio spectra, (Mart\'\i{} 1993;
Mirabel 1994) similar to Cyg X-1 and GX 339-4.  In addition, both have
associated weak arcmin-scale radio lobes (Mirabel 1994) providing
direct evidence for the action of a jet on the ISM in the past, if not
at present.  LMC X-1 and LMC X-3 are very luminous X-ray sources from
which radio emission has never been detected despite several deep
observations (Fender et al. 1998). Table 1 summarises the observed
radio flux densities and distance estimates of the persistent BHC
systems.

\subsection{Z-type X-ray binaries}
\label{subsec:Z}

The six Z-type X-ray binaries (Z-sources) are believed to be low
magnetic field neutron stars accreting at, or just below, the
Eddington limit, and as such are amongst the most luminous persistent
X-ray sources in our Galaxy (e.g. van der Klis 1995, 1999). Their name
arises from a characteristic pattern they trace out in X-ray
colour-colour and hardness-intensity diagrams and they are further
characterised by their X-ray timing properties (e.g. van der Klis
1995, 1999). Penninx (1989) reported that four of the six Z-sources
had been detected at radio wavelengths, and suggested that they all
had approximately the same radio luminosity when on the `Horizontal
Branch' of the X-ray colour-colour or hardness-intensity diagrams --
see e.g. van der Klis (1995).  Following his suggestion, the remaining
two Z-sources were detected at the predicted levels (Cooke \& Ponman
1991; Penninx et al. 1993).

These systems are usually, but not always, detected by radio
observations with a sensitivity of $\geq 0.1$ mJy.  As noted above,
the most persistent radio emission is associated with the `Horizontal
Branch' of the X-ray emission; weaker emission appears to be
associated with the `Normal Branch', and radio emission appears to be
`off' during the `Flaring Branch' (Hjellming \& Han 1995 and
references therein).  They show generally flat radio spectra except
during occasional flaring events (most prominent in Sco X-1) during
which time an optically thin component is superposed. Bradshaw,
Geldzahler \& Fomalont (1997) have found evidence for periodic flux
density variations from Sco X-1, and more recently Bradshaw, Fomalont
\& Geldzahler (1999) have established the distance and verified the
existence of radio-emitting outflows from the system. The radio flux
densities and distance estimates of the Z-sources are summarised in
table 2.

\subsection{Atoll-type X-ray binaries}
\label{subsec:Atoll}

The Atoll-type X-ray binaries (`Atoll sources'), like the Z sources,
are low mass X-ray binaries containing low magnetic field accreting
neutron stars. However, they are believed to be accreting at around an
order of magnitude lower rate than the Z sources (see e.g. van der
Klis 1995).  As discussed above, while vP95 listed only 11 Atoll
sources, it now seems likely that the majority of other low-luminosity
X-ray binaries classified as Bursters and/or Dippers (and/or
occasionally Transient) are also Atoll sources. Only a very small
number of these systems have reported detections at radio wavelengths
(Hjellming \& Han 1995). Grindlay \& Seaquist (1986) report the
detection of a weak ($0.49 \pm 0.12$ mJy) radio signal from 4U
1820-30, which has not been confirmed.  Mart\'\i{} et al. (1998) report the
possible detection of variable radio emission from GX 354+0 which
peaks at a level of $\sim 0.5$ mJy but which is undetected to $\leq
0.3$ mJy in the majority of their observations. Gaensler, Stappers \&
Getts (1999) also report a transient radio detection of the Atoll-like
millisecond X-ray pulsar SAX 1808.4-3658.  The only convincing and
repeated detection of persistent radio emission from an Atoll source
is that of GX 13+1, an unusual system sharing some of the properties
of both Atoll and Z sources (Penninx 1990; Homan et al. 1998).

As already stated, the majority of other, unclassified X-ray
binaries are also likely to be Atoll sources. The X-ray binary radio
surveys of Nelson \& Spencer (1988 -- northern hemisphere) and McKie
(1997; see also Spencer et al. 1997 -- southern hemisphere) typically
reached flux density detection limits of 2.0 and 0.2 mJy
respectively. Without doubt some of the `miscellaneous' X-ray binaries
observed (but not detected)
were unclassified low-luminosity Atoll sources; none were
detected as radio sources.  In addition, some neutron-star transient
systems (e.g. Aql X-1) are classified as Atoll-like and have been
observed to produce radio emission during outbursts (presumably at
high accretion rates and/or during state transitions).  The radio flux
densities and distance estimates of the persistent Atoll-sources are
summarised in table 3.

\subsection{X-ray pulsars}
\label{subsec:XRP}

More than 30 X-ray pulsar systems, containing a high magnetic field
neutron star which disrupts the inner accretion disc and channels the
accretion flow onto its magnetic poles (White, Nagase \& Parmar 1995;
Bildsten et al. 1997), are known. Many of these systems are in high
mass X-ray binaries, suggesting that the neutron stars are relatively
young. Not one of the high-field X-ray pulsars has ever been convincingly
detected as a synchrotron radio source (Fender et al. 1997). Mart\'\i{} et al.
(1997) report a marginal radio detection of the X-ray pulsar system GX
1+4 but this has yet to be confirmed, and in any case may be
consistent with thermal free-free emission from the red giant wind in
this system. Fender et al. (1997) showed that there was a significant
anticorrelation between the properties of radio emission and X-ray
pulsations from X-ray binaries, which they suggested arose because of
the disruption of the inner accretion disc preventing the formation of
an outflow from the system.

Table 4 lists ten persistent X-ray pulsar systems for which there are
good limits to the radio flux density
and reasonable distance estimates. None
of the systems are detected as radio sources. It is worth noting that
(at least) an additional six {\em transient} X-ray pulsars (4U
0115+63, GS 0834-430, A 0535-66, A 1118-616, 4U 1145-619, 4U 1416-62)
have not been detected to comparable limits (Fender et al. 1997; McKie
1997; some distance estimates in Negueruela 1998).

As noted above, there is a reported detection of transient radio
emission from the accretion-powered millisecond X-ray pulsar SAX
1808.4-3658 (Gaensler et al. 1999). However, in nearly all its
properties this system is more like an Atoll source than an X-ray
pulsar, due to the low ($\leq 10^9$G, c.f. $\geq 10^{11}$G for other
X-ray pulsars) magnetic field.

\begin{figure*}
\centering
\leavevmode\epsfig{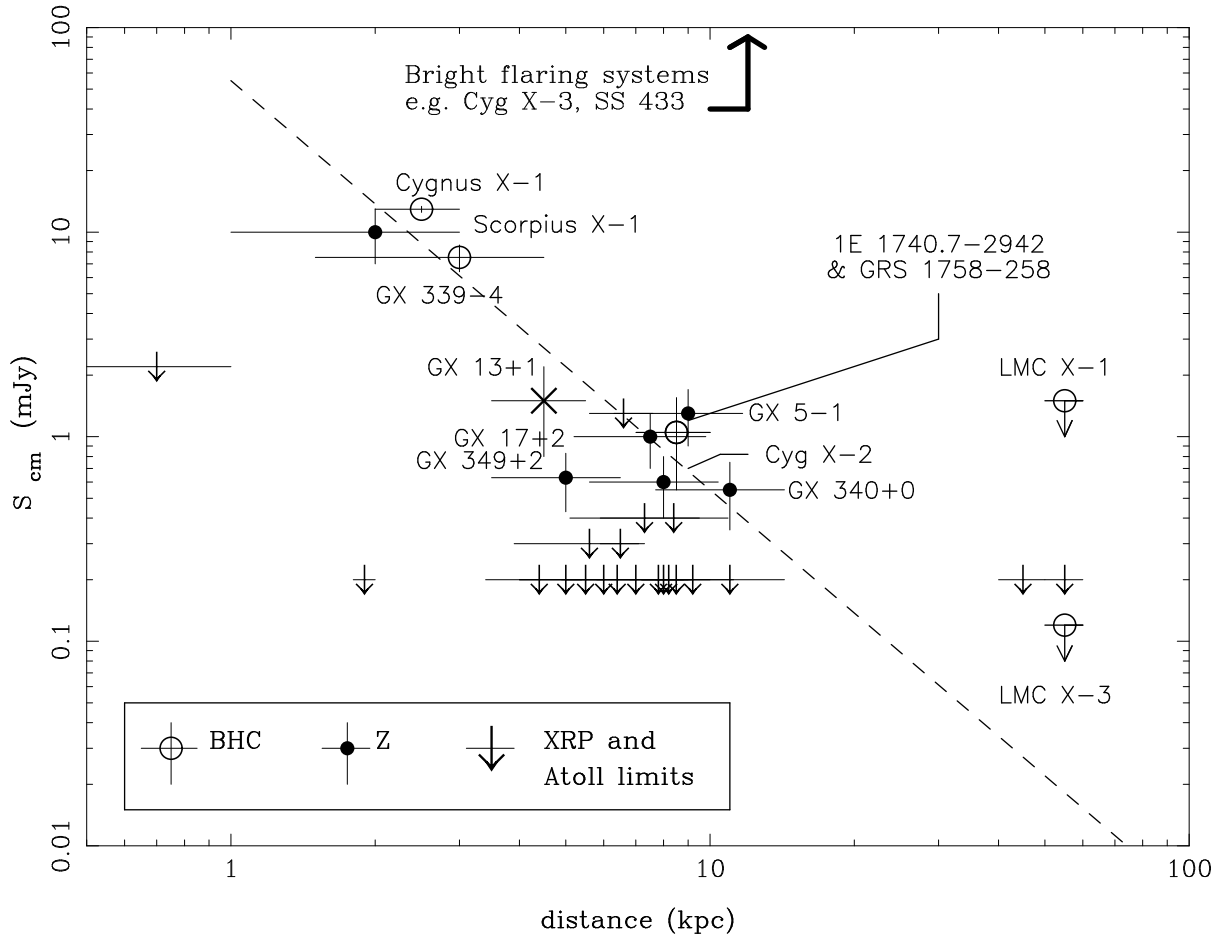}
\caption{A plot of the mean radio emission against distance estimate
for the persistent BHC and Z-source X-ray binaries, using the values
listed in Tables 1 and 2. In addition the upper limits to radio
emission from the Atoll (except GX 13+1, see text) and X-ray pulsar
systems are indicated, using the values in Tables 3 \& 4.  A single
function, described in the text, and indicated on the figure by the
dashed line, can be reasonably fitted to all the persistent BHCs and Z
sources, implying a similar centimetric radio luminosity for all
the sources.}
\label{fig:relation}
\end{figure*}

\subsection{Other, peculiar, systems}

In addition to the classes of system described above, there are
several systems which are persistently detected by X-ray missions but
which are not easily classified. These include Cyg X-3, Cir X-1 and,
for the past 5 years, GRS 1915+105, although this last source was
almost certainly `off' prior to 1994. Several of these systems,
notably those mentioned above, are bright and variable radio
sources. However, while the quiescent state of their radio emission is
relatively weak with a flat spectrum, their radio emission is often
dominated by the superposition of many components which evolve from
optically thick to optically thin. It seems that radio jet production
in these systems is more sporadic and violent than in the persistent
sources, and it is not within the scope of this paper to try and
identify in detail areas of common astrophysics between the production
of radio emission in persistent and transient sources.

\section{A common radio luminosity for persistent BHC and Z sources}
\label{sec:common}

Fig 1 plots the mean radio flux densities and best distance estimates
for the persistent BHC and Z sources, plus the anomalous Atoll source
GX 13+1, from the data listed in Tables 1-4.  The limits on radio
emission from the other Atoll sources and the persistent X-ray pulsar
systems are also indicated. The weakness of the radio emission from
the X-ray pulsar and Atoll sources compared to the persistent BHC and Z
sources is immediately apparent. In fact the data suggest a common
mean radio luminosity for the persistent BHC and Z sources, which is far
above that of the other persistent X-ray binaries. It is worth
stressing that this radio luminosity is still orders of magnitude
below that of the more extreme and poorly classified systems such as
Cyg X-3, SS 433 etc.

In order to test the goodness of fit of a common observed radio
luminosity for all the sources, a straight line was fitted to the
relationship between the (base ten) logarithm of centimetre radio
flux density, $S_{\rm cm}$ (in mJy) and estimated distance, $d$ (in kpc).
Both the intercept and gradient were treated as free parameters and
our least squares fitting procedure allowed for errors on both flux
and distance -- the appropriate errors on the log quantities having
been derived from the tabulated flux and distance errors using
Monte Carlo simulations. We included only the ten galactic sources
in our fits, since the two LMC sources had only upper limits on the
observed flux. The resulting best fit relation was

\begin{equation}
\log_{10} S_{\rm cm}
= (2.03 \pm 0.40) \quad - \quad (2.46 \pm 0.51) \, \log_{10} d
\label{eq:free_slope}
\end{equation}

\noindent
($\chi^2_{\rm red} = 0.514$)
which is clearly consistent with a $d^{-2}$ relation and therefore
a common luminosity. Fixing the slope of the relation to be equal to
$-2$ produced a similar fit, although now with a smaller error on
the intercept:

\begin{equation}
\log_{10} S_{\rm cm}
= ( 1.71 \pm 0.10) \quad - \quad 2 \, \log_{10} d
\label{eq:fixed_slope}
\end{equation}

\noindent
($\chi^2_{\rm red} = 0.527$)
which is shown by the dotted line on Fig \ref{fig:relation}. Expressed in 
terms of flux and distance (in kpc) this relation corresponds to

\begin{equation}
S_{\rm cm} = \frac{55 \pm 13}{d^2} \phantom{00} {\rm mJy}
\label{eq:fixed_flux}
\end{equation}

A possible explanation for
the low values of $\chi^2_{\rm red}$ may be that some of the flux and
distance errors tabulated in Tables 1 \& 2 have been over-estimated: if
the quoted errors -- particularly those in distance -- were correct
then the data would be unlikely to lie so close to the best fit
straight line shown in Fig \ref{fig:relation}.  We discuss this point
further below, but remark for the moment that the agreement of the
data with an inverse square relation between flux and distance is
clearly convincing. This is a surprising result given that both
different accretion structures and outflow velocities (see below) in
the neutron star and black hole systems might be expected.

For a flat spectrum from 30 to 2 cm (1 to 15 GHz), this corresponds to
an observed radio synchrotron luminosity of $10^{30}$ erg s$^{-1}$
($10^{23}$ W) for these sources. It should be stressed however that a
high-frequency cut-off in synchrotron emission has yet to be found in
any of these sources, and the total (ie. radio -- mm -- infrared)
synchrotron luminosity is likely to be orders of magnitude higher (see
for example Fender et al. 1997c and Mirabel et al. 1998 where a
synchrotron luminosity $\geq 10^{36}$ erg s$^{-1}$ is inferred for GRS
1915+105 from the observation of the flat spectral component to 2
$\mu$m).

\subsection{The sample and selection effects}
\label{sec:selection}

As discussed in the introduction, the sources under discussion, the
neutron star Z-sources and persistent BHCs, are unique in their
relatively steady, bright X-ray emission, implying steady accretion at
high luminosity. Given the number and sensitivity of X-ray missions
over the past 30+ yr, the sample is likely to be more or less complete
for the entire Galaxy, LMC and SMC.

The sample is clearly dominated by sources in the vicinity of the
Galactic centre, with distances between 7 -- 10 kpc. This clustering
of data points is suggestive of selection effects dominating the fit
to the data, as a result of small numbers in the sample. However, as
discussed above, the sample is effectively volume-limited and cannot
be expanded. One selection effect which could significantly bias the
fit would be if the Galactic centre systems were being detected only
when they came up above the detection limits of typical radio
observations. In this case, a mean flux density calculated from an
unrepresentative sample of positive detections could be very similar
for all the sources, regardless of their true mean radio
flux. However, this does {\em not} appear to be the case :
comprehensive high-sensitivity observations of Cyg X-2 and GX 17+2
(Hjellming et al. 1990; Penninx et al. 1988), among the most distant
sources in the sample, consistently detect these sources and
accurately measure the mean flux density.

\section{Beaming and the luminosity function}
\label{sec:is_persistent?}

It has been speculated that low-level radio emission from
the persistent BHC and Z-source X-ray binaries could arise in a
continuous jet (e.g. Hjellming \& Han 1995). It has also
been suggested that the velocities of jets from accretion discs
should approximately reflect the escape velocity of the central
object, i.e. jets from black holes will have velocities
$\geq 0.9c$, those from neutron stars $\sim 0.3c$ etc. (e.g.
Livio 1997).

Combining these ideas, a scenario can be envisaged whereby the
low-level radio emission from persistent BHCs and Z-sources originates
in compact jets of velocities $\sim 0.9c$ and $\sim 0.3c$
respectively.  In the light of our result that all persistent BHC and
Z-source X-ray binaries have the same mean radio luminosity at
centimetre wavelengths, however, there are problems with this
interpretation, based upon the Doppler boosting associated with a
relativistic jet.

Relativistic jets are significantly Doppler boosted; for a continuous
jet at a given angle to the line of sight, $\theta$, the rest-frame
flux of the source, $S$, is boosted to an apparent value,

\begin{equation}
S^{\prime} = S D^{2-\alpha}
\label{eq:s_prime}
\end{equation}

where $\alpha$ is the spectral
index of the radio emission, and the relativistic Doppler factor $D$
is defined as

\begin{equation}
D = [ \gamma (1\mp\beta \cos \theta)]^{-1}
\label{eq:boost_factor}
\end{equation}

($\mp$ for approaching and receding components respectively) and
$\gamma$ is the Lorentz factor

\begin{equation}
\gamma = (1-\beta^2)^{-1/2}
\label{eq:gamma}
\end{equation}

where $\beta$ is the velocity of the jet expressed as a fraction of
$c$.  As an example, at an angle to the line of sight of 30 degrees, a
flat-spectrum ($\alpha = 0$) symmetric jet of velocity $0.9c$ would
appear 2.4 times brighter than the same jet at a velocity of $0.3c$,
and 3.9 times brighter than in the rest frame.  The total flux
observed will be a sum of the approaching and receding jets. Note that
for jets near to the plane of the sky {\em both} approaching and
receding jets can be de-boosted (as is the case for GRS 1915+105;
Mirabel \& Rodriguez 1994; Fender et al. 1999b). Figure 2 illustrates
the ratio of observed to intrinsic flux expected from symmetric jets
at $0.3c$ and $0.9c$ for all inclinations.

Assuming naively that any jet is approximately perpendicular to the
orbital plane of the system, then the angle, $\theta$, of the jet to
the line of sight would be equal to the inclination, $i$, of the
orbital plane to the line of sight.  Estimates of the orbital
inclinations of the systems in question are very limited. A survey of
the literature does not reveal any particular bias in inclination
estimates for any of the classes of source, so we will assume that the
inclinations are uniformly distributed in $\cos i$. It certainly appears
likely that the systems show a significant spread in inclination. This
immediately presents a problem for the hypothesis that the
radio emission from these systems originates in Doppler boosted
jets. The problem may be stated qualitatively as follows.

If the ten galactic systems are viewed at a range of different
inclinations, their {\em observed\/} fluxes will be boosted by a
range of different Doppler factors. Thus, if their observed fluxes
obey the inverse-square relationship with distance given by Eq.
(\ref{eq:fixed_flux}),
their {\em intrinsic\/} fluxes will {\em{not}\/} in general
obey this relationship. It would, therefore, seem unlikely that one
should obtain such a relation between distance and observed flux by
chance, since it would require a series of fortunate coincidences
in order that the intrinsic fluxes and inclinations yield observed
fluxes in agreement with the fitted relation.

One resolution of this problem would be if all the systems were
observed at approximately the same inclination, since the sources
would then all be Doppler boosted by the same factor. As we have
already remarked, however, this possibility appears incompatible
with the inclination estimates reported in the literature. Even if
we choose to regard these estimates as unreliable, it seems
reasonable to suppose that the orbital planes of the ten systems
should be randomly sampled from a uniform distribution over all
possible orientations. It is then straightforward to show that the
probability of drawing a sample of ten sources, with inclinations
all lying within, say, an interval of 5 degrees, is less than
$10^{-10}$.

Is it possible that the inclinations of the observed systems are {\em
selected\/} to lie within a narrow range? One plausible mechanism for
such a selection effect might be if their intrinsic fluxes were too
faint to be detected, but there exists a critical inclination at which
the Doppler boosting factor is sufficient to raise the observed fluxes
above the detection limits of typical radio observations. As remarked
above, however, our sample is {\em not\/} dominated by systems whose
flux lies close to the detection limit. The observed fluxes listed in
Tables 1 \& 2 span a range of more than a factor of 20, arguing in a
favour of a wide range of different inclinations consistent with the
estimated limits.

\subsection{Monte Carlo simulations}

We have used extensive Monte Carlo simulations to investigate the
effect of beaming and the width of the intrinsic radio luminosity
function (LF) of the persistent BHC and Z sources. The width is
defined such that the LF is uniform in $\log L$ between $\log
(L / width)$ and $\log (L \times width)$. We have defined a {\em critical
width} for which, after running $10^4$ simulations, 90\% of the sample
have $\chi^2_{\rm red} > 3$. Not surprisingly, the higher values of
$\beta$ we consider for the jets, the narrower the intrinsic LF must
be in order to produce the observed inverse square relationship. The
results of the simulations are listed in Table 5. In addition to
considering the same value of $\beta$ for both jets from black hole
and neutron star systems, we also consider what may be considered the
`canonical' model, where jets from black hole systems have $\beta =
0.9$ and jets from neutron star systems have $\beta = 0.3$.

\begin{figure}
\label{fig:boost}
\centering
\leavevmode\epsfig{file=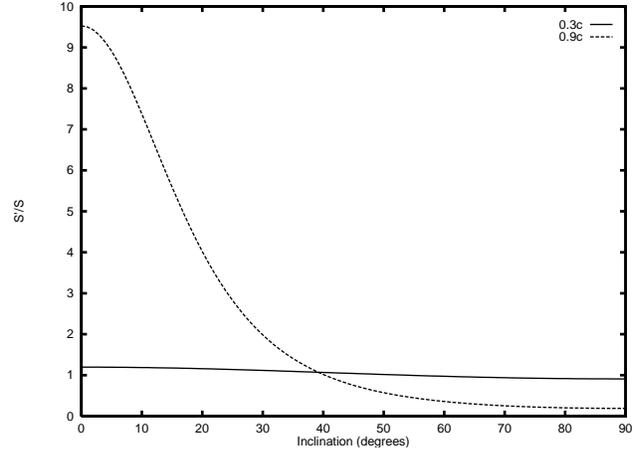,angle=270,width=8.5cm,clip}
\caption{Effects of Doppler boosting for symmetric ejections of 0.3
and 0.9 c, as envisaged for neutron star and black hole systems
respectively, and with a spectral index of 0 (as observed for most
systems outside of flaring events).  $S^{\prime} / S$ is the sum of
both approaching and receding jets normalised to the emission in the rest
frame. While the Doppler boosting associated with a velocity of 0.3 c
is quite small ($S^{\prime} / S$ = 0.9 -- 1.2) that associated with a
velocity of 0.9 c can be very great (0.2 -- 9.5).}
\end{figure}

\begin{table}
\centering
\begin{tabular}{ccc}
\hline
$\beta$ & $\beta$ & Critical width of \\
 (BHC)  &  (Z)    & Luminosity Function \\
\hline
0.0 & 0.0 & 26.1 \\
\hline
0.5 & 0.0 & 26.0 \\
    & 0.1 & 25.8 \\
    & 0.3 & 25.7 \\
    & 0.5 & 25.5 \\
\hline
0.7 & 0.0 & 24.5 \\
    & 0.1 & 24.4 \\
    & 0.3 & 24.3 \\
    & 0.5 & 24.0 \\
    & 0.7 & 23.5 \\
\hline
0.9 & 0.0 & 22.9 \\
    & 0.1 & 22.7 \\
    & 0.3 & 22.0 \\
    & 0.5 & 21.5 \\
    & 0.7 & 21.0\\
    & 0.9 & 15.0 \\
\hline
\end{tabular}
\caption{Critical width of radio luminosity function for different
values of $\beta$ $(=v/c)$ for the population of persistent BHC and Z source
X-ray binaries}
\end{table}

It is clear from the results of the MC simulations presented in table
5 that for jet velocities in the range 0 -- 0.7$c$ the intrinsic
radio luminosity of the BHC and Z-sources must be the same to within a
factor of 25 or so. This is a result of the relative unimportance of
Doppler boosting at these velocities. For higher jet velocities the
intrinsic luminosities need to be even closer together: for jets in
both types of systems with velocities of 0.9$c$ they must be
intrinsically within a factor of 15 in radio luminosity in order to
produce the observed $d^{-2}$ relation. Thus we can conclude that the
classes of BHC (in the low/hard X-ray state) and Z sources have a
common intrinsic radio luminosity within an order of magnitude or so.

We can also use the MC simulations to quantify how much weaker radio
sources the Atoll and X-ray pulsar systems are, as a class, compared
to the BHC and Z-sources. We use the observed upper limits on radio
emission and distance estimates given in tables 3 \& 4. The maximum
mean luminosities of these two classes, compared to that obtained for
the combined BHC and Z-sources, are given in table 6. From this we can
assert that, as a class, the mean radio luminosity of Atoll sources is
more than a factor of five below that of the BHC and Z sources. For
the X-ray pulsar systems the limits are even stronger; they are at
least an order of magnitude fainter, as radio sources, than the BHC
and Z sources.

\begin{table*}
\centering
\begin{tabular}{cccccc}
\hline
Source & Compact & $S_{\nu} / (\rm{kpc}^2)$ & \multicolumn{3}{c}{Inferred
physical characteristics}\\
type   & object  & (mJy) & accretion rate ($\dot{m}_{\rm Edd})$ & magnetic field
(B) & inner disc radius (km) \\
\hline
BHC (low/hard state) & BH & $55 \pm 13$ & $\leq 0.1$ & -- & few $\times 100$ \\
Z (horizontal branch) & NS &  & 0.1 -- 1.0 & $10^9$--$10^{10}$
& few $\times 10$ \\
\hline
Atoll & NS & $\leq 10.0 \pm 2.4$ & 0.01 -- 0.1 & $10^9$--$10^{10}$ & few $\times$ 10 \\
X-ray pulsar & NS & $\leq 6.6 \pm 2.4$ & $\leq 1.0$ & $\geq 10^{12}$ & $\geq 1000$ \\
\hline
\end{tabular}
\caption{
Comparison of derived mean intrinsic radio luminosities for the BHC/Z,
Atoll and X-ray pulsar classes of X-ray binary, plus simple
interpretations of their physical differences}
\end{table*}

\section{Discussion}
\label{sec:discussion}

We have found that the BHC and Z-source X-ray binaries share a common
mean radio luminosity to within a factor of 15--25, depending on the
velocity of the inferred outflows. The Atoll sources are $\geq 5$
times fainter; the X-ray pulsar systems $\geq 10$ times so. One reason
that these results are surprising is the different inferred accretion
modes for the BHCs and Z-sources. In BHCs in the low/hard X-ray state
the `standard' (thin, cold, optically thick) accretion disc is
believed to be truncated many Schwarzchild radii from the central
black hole and replaced in the inner regions by an optically thin,
radiatively inefficient quasi-spherical flow (Advection-dominated
accretion flows; see Svensson 1998 and references therein). However in
Z and Atoll sources the `standard' accretion disc is believed to reach
to almost the surface of the neutron star (e.g. van der Klis 1999).

\subsection{Jets ?}

We have established that Doppler boosting is unlikely to affect the
observed radio luminosities of the BHC and Z sources by more than an
order of magnitude. Thus, assuming the emission is incoherent, we can
apply the limiting brightness temperature of $10^{12}$ K which results
from second-order inverse Compton losses. As a result we find that the
emission at 2 GHz must arise in a region $\geq 10^{12}$ cm (for a
spherical emitting region). This is a significant size scale, larger
than the inferred binary separations of most, maybe all, of the systems 
(e.g. $\sim 3 \times 10^{11}$ cm for
the Z source Sco X-1). A cone of opening half-angle 10
degrees in the plane of the sky would require a length of $10^{13}$ cm
to produce the same observed surface area; angling the jet more
towards the line of sight or making the opening angle smaller only
increases this dimension. Similarly, attributing at least some of the
observed radio flux to optically thin emission, less efficient than
optically thick, also increasing the required emitting volume.
Coupled with the recent imaging of collimated outflows from both Sco
X-1 (Bradshaw et al. 1999) and Cyg X-1 (Stirling et al. 1998; de la
Force et al. in prep) the observational evidence seems to point to
extended radio-emitting outflows in all BHCs, Z sources and GX
13+1.

\subsection{Why not Atoll sources ?}

Why are Z sources so much brighter radio emitters than the Atoll
sources ? the inferred differences between the two classes are a
stronger magnetic field and higher accretion rate in the Z sources. We
do not believe that the magnetic field plays much of a role in this
difference:

\begin{itemize}
\item{The inferred magnetic field in Z sources lie between those of
the Atoll sources and the X-ray pulsars, both of which we shown to be
significantly less luminous radio sources.}
\item{The presence of kilohertz quasi-periodic oscillations in both Z
and Atoll sources (van der Klis 1999 and references therein) implies
that the accretion flow in both classes is not truncated by the
magnetic field and instead reached almost to the surface of the
neutron star.}
\end{itemize}

Instead, it seems likely that it is the accretion rate which is the
origin of the difference. This is supported by the radio detections of
the Atoll source GX 13+1 at a similar level to the Z sources; this
system is believed to be accreting at a higher rate ($\sim 10^{17}$ g
s$^{-1}$ $\equiv 0.1$ Eddington) than the other Atoll sources. In
addition, we can imagine that the occasional detections of Atoll-type
sources at radio wavelengths are associated with transient periods of
high accretion rates comparable to those continuously occuring in the
Z sources.

Alternatively, or perhaps additionally, transient radio emission seems to be
produced at points of change in the X-ray `state' of a system. Perhaps
GX 13+1 and the Z sources change X-ray `state' more often, or
physically in a more dramatic way, than the other Atoll sources, and
hence are more prone to significant mass ejections.

\subsection{Why not X-ray pulsars ?}

As mentioned in section 2.4, no X-ray pulsar has ever been detected as
a synchrotron radio source. As originally suggested by Fender et
al. (1997b) we believe this is due to the truncation of the inner
accretion flow by strong neutron star magnetic fields which force the
accreting material to flow along the field lines towards the magetic
poles. We can now definitively state that the X-ray pulsars are at
least one order of magnitude fainter than the BHC and Z sources as a
population of radio emitters. 

The apparent exception to this rule is the recent detection of
transient radio emission from SAX 1808.4-3658 (Gaensler et
al. 1999). In fact, this observations seems to confirm, rather than
violate the above hypothesis, as in SAX 1808.4-3658 the magnetic field
appears to be so weak ($\leq 10^9$G) as to allow the nearly-Keplerian
flow of material almost to the neutron star surface (Wijnands \& van
der Klis 1998). In this case the source is in nearly all respects
Atoll-like and the detection of transient, weak, radio emission is
consistent with this picture. It appears that somewhere in the range
$10^9$ -- $10^{11}$ G, the magnetic field of a neutron star becomes so
strong that its affect on the inner disc structure is enough to
prevent the formation of a radio-emitting outflow.

\section{Conclusions}

We have investigated the radio detections and upper limits on the
radio emission from persistent (i.e. non transient) X-ray
binaries. Whilst always bearing in mind that the sample is not large,
our conclusions are summarised in Table 6, and are :

\begin{itemize}
\item{The BHCs (in the low/hard state) and the Z sources (on the
horizontal branch) share a common mean observed radio luminosity corresponding
to $(55 \pm 13) / d^2$ mJy, where $d$ is the distance to the source in
kpc.}
\item{Depending on the degree of Doppler boosting of the radio
emission, this implies a common intrinsic radio luminosity to within a
factor of 25 (decreasing as Doppler boosting becomes more important to
e.g. a factor of 15 if both BHCs and Z sources have jets with $v =
0.9c$).}
\item{Upper limits on radio emission from Atoll and X-ray pulsar
populations as a whole show that they are in general at least 5 and 10
times fainter, respectively, than the BHC/Z systems.}
\item{Assuming that the radio emission from BHC/Z systems arised in
jets for which Doppler boosting is not very significant, we find that
all these systems are likely to be generating radio-emitting outflows
or jets whose physical scales are significantly larger than the
binary orbits.}
\end{itemize}

Combining these results with knowledge of the nature of accretion in
different types of X-ray binaries, we can surmise that the following
physical conditions are required for formation of a radio jet :

\begin{itemize}
\item{A dipole magnetic field of $\leq 10^{10}$ G associated with the
accreting compact object, allowing the formation of an accretion flow
to $\leq 1000$ km which is not channeled onto the magnetic poles of
the neutron star.}
\item{A high accretion rate ($\geq 0.1$ Eddington) and/or dramatic
physical changes in the accretion mode which result in the ejection of
disc material.}
\end{itemize}

and further that the coupling between accretion and outflow in
persistent systems (excluding X-ray pulsars) is comparable for both
neutron stars and black holes, and therefore probably does not require
the presence of either a surface or an event horizon.
 
Observation of exactly what causes the Atoll
sources to occasionally produce radio emission, and determination of
the high-frequency spectrum of the radio emission from the Z sources
(to see if they, like the BHCs, possess a flat spectrum through mm
wavelengths) are amongst the many important future observations to be
made in this field.

\section*{acknowledgements}

RPF wishes to thank Eric Ford, Guy Pooley, Michiel van der Klis, Jan van
Paradijs and Ben Stappers for useful discussions. This work was
supported in part by ASTRON grant 781-76-017, and in part by EC Marie
Curie Fellowship ERBFMBICT 972436.

\appendix


\begin{thebibliography}{}

\bibitem{}
Anantharamaiah K.S., Dwarakanath K.S., Morris D., Goss W.M.,
Radhakrishnan V., 1993, ApJ, 410, 110

\bibitem[]{}
Berendsen S.G.H., Fender R., Kuulkers E., Heise J., van der Klis M., 
1999, MNRAS, submitted

\bibitem[]{}
Bildsten L. et al., 1997, ApJS, 113, 367

\bibitem{}
Bradshaw C.F., Geldzahler B.J., Fomalont E.B., 1997, ApJ, 481, 489

\bibitem[]{}
Bradshaw C.F., Fomalont E.B., Geldzahler B.J., 1999, ApJ, 512, L121

\bibitem[]{}
Brocksopp C., Fender R.P., Larionov V., Lyuty V.M., Tarasov A.E.,
Pooley G.G., Paciesas W.S., Roche P., 1999, MNRAS,  309, 1063

\bibitem{}
Callanan P.J., Charles P.A., Honey W.B., Thorstensen J.R., 1992,
MNRAS, 259, 295

\bibitem{}
Casares J., Charles P.A., Kuulkers E., 1998, ApJ, 493, L39

\bibitem[]{}
Chakrabarty D., 1998, ApJ, 492, 342

\bibitem[]{}
Christian D.J., Swank J.H., 1997, ApJS, 109, 177

\bibitem{}
Cooke P.A., Ponman T.J., 1991, A\&A, 244, 358

\bibitem{}
Cowley A.P., Crampton D., Hutchings J., 1979, ApJ, 207, 907

\bibitem[]{}
Corbel S., Fender R.P., Durouchoux P., Sood R.K., Tzioumis A.K.,
Spencer R.E., Campbell-Wilson D., 1997, In: Proceedings of the 4th
Compton Symposium, eds. Dermer C.D., Strickman M.S., Kurfess J.D., AIP
conf. proc. 410, Woodbury, New York, p. 937

\bibitem[]{}
Corbel S., Fender R.P., Tzioumis A.K., Nowak M., McIntyre V.,
Durouchoux P., Sood R., 2000, A\&A, submitted

\bibitem{}
Crampton D., Cowley A.P., Hutchings J.B., Kaat C., 1976, ApJ, 207, 907

\bibitem{}
Falcke H., Biermann P.L., 1996, A\&A, 308, 321

\bibitem[]{} 
Fender R.P., 2000, In: `Astrophysics and Cosmology : A collection of
critical thoughts', Eds. Kundt W., van den Bruck C., Springer Lecture
Notes in Physics, in press ({\bf astro-ph/9907050})

\bibitem{}
Fender R.P., Bell~Burnell S.J., Waltman E.B., 1997, Vistas Astron.,
41, 3


\bibitem{}
Fender R.P., Spencer R.E., Newell S.J., Tzioumis A.K., 1997a, MNRAS,
286, L29

\bibitem{}
Fender R.P., Roche P., Pooley G.G., Chakrabarty D., Tzioumis A.K.,
Hendry M.A., Spencer R.E., 1997b, in Winkler C., Courvoiser T.J-.L.,
Durouchoux P., eds., {\em Proc. 2nd INTEGRAL workshop : The
Transparent Universe}, ESA SP-382, 303

\bibitem{}
Fender R.P., Pooley G.G., Brocksopp C., Newell S.J., 1997c, MNRAS,
290, L65

\bibitem{}
Fender R.P., Southwell K., Tzioumis A.K., 1998, MNRAS, 298, 692

\bibitem[]{}
Fender R.P. et al. 1999a, ApJ, 519, L165

\bibitem[]{} 
Fender R.P., Garrington S.T., McKay D.J ., Muxlow T.W.B., Pooley G.G.,
Spencer R.E., Stirling A.M. \& Waltman E.B. 1999b, MNRAS, 304, 865

\bibitem[]{}
Fender R.P., Pooley G.G., Durouchoux P., Tilanus R.P.J., Brocksopp C.,
2000, MNRAS, in press

\bibitem[]{}
Gaensler B.M., Stappers B.W., Getts T.J., 1999, ApJ, 522, L117

\bibitem{}
Gies D.R., Bolton C.T., 1986, ApJ, 304, 371

\bibitem[]{}
Grindlay J.E., Seaquist E.R., 1986, ApJ, 310, 172

\bibitem{}
Han X.H., 1993, PhD thesis, NMIMT, New Mexico

\bibitem[]{}
Hannikainen D.C., Hunstead R.W., Campbell-Wilson D., Sood R.K., 1998,
A\&A, 337, 460

\bibitem{}
Hasinger G., van der Klis M., 1989, A\&A, 225, 79

\bibitem{}
Hjellming R.M., Han X.H., 1995, in Lewin W.H.G., van Paradijs J., van
den Heuvel E.P.J., eds., {\em X-ray binaries}, CUP, p. 308


\bibitem{}
Hjellming R.M., Johnston K.J., 1988, ApJ, 328, 600

\bibitem{}
Hjellming R.M., Rupen M.P., 1995, Nat, 375, 464

\bibitem{}
Hjellming R.M., Han X.H., Cordova F.A., Hasinger G., 1990, A\&A, 235, 147

\bibitem{}
Homan J., van der Klis M., Wijnands R., Vaughan B., Kuulkers E., 1998,
ApJ, 499, L41

\bibitem{}
Ilovaisky S.A., Chevalier C., Motch C., 1982, A\&A, 114, L71

\bibitem{}
Kaper L., van Loon J. Th., Augusteijn T., Goudfrooij P., Patat F.,
Waters L.B.F.M., Zijlstra A.A., ApJ, 475, L37

\bibitem{}
Kuulkers E., van der Klis M., Oosterbroek T., Asai K., Dotani T., van
Paradijs J., Lewin W.H.G., 1994, A\&A, 289, 795

\bibitem{}
Livio M., 1997, in Wickramasinghe D.T., Ferrario L., Bicknell G.V.,
eds, IAU Coll. 163, Accretion phenomena and related outflows, ASP
conf. ser. vol. 121, p.845.

\bibitem{}
Mart\'\i{} J., 1993, PhD thesis, University of Barcelona

\bibitem{}
Mart\'\i{} J., Rodriguez L.F., Mirabel I.F., Paredes J.M., 1996,
A\&A, 306, 449

\bibitem{}
Mart\'\i{} J., Mirabel I.F., Chaty S., Rodriguez L.F., 1997,
in Winkler C., Courvoiser T.J-.L.,
Durouchoux P., eds., {\em Proc. 2nd INTEGRAL workshop : The
Transparent Universe}, ESA SP-382, 303

\bibitem{}
Mart\'\i{} J., Mirabel I.F., Rodriguez L.F., Chaty S., 1998, A\&A,
332, L45

\bibitem[]{}
McKie S., 1997,  A Southern Survey for radio emitting X-ray binaries, MSc
thesis, University of Manchester

\bibitem[]{}
Mendez M., van der Klis M., 1997, ApJ, 479, 926

\bibitem{}
Mirabel I.F., 1994, ApJS, 92, 369

\bibitem{}
Mirabel I.F., Rodriguez L.F., 1994, Nature, 371, 46

\bibitem{}
Mirabel I.F., Dhawan V., Chaty S., Rodriguez L.F., Mart\'\i{} J., Robinson
C.R., Swank J., Geballe T.R., 1998, A\&A, 330, L9

\bibitem{}
Mirabel I.F., Rodriguez L.F., 1999, ARA\&A, 37, 409

\bibitem{}
Negueruela I., 1998, A\&A 339, 505

\bibitem{}
Nelson R.F., Spencer R.E., 1988, MNRAS, 234, 1105

\bibitem{}
Pooley G.G., Fender R.P., Brocksopp C., 1999, MNRAS, 302, L1

\bibitem{}
Penninx W., Lewin W.H.G., Zijlstra A.A., Mitsuda K., van Paradijs J.,
van der Klis M., 1988, Nature, 336, 146

\bibitem{}
Penninx W., 1989, in Hunt J. and Battrick B., eds., {\em 23rd ESLAB
Symp. on Two Topics in X-ray Astronomy}, Bologna, Italy, ESA SP-296, p.185

\bibitem[]{}
Penninx W.H., 1990, PhD thesis, University of Amsterdam

\bibitem{}
Penninx W., Zwarthoed G.A.A., van Paradijs J., van der Klis M., Lewin
W.H.G., Dotani T., 1993, A\&A, 267, 92

\bibitem{}
Reynolds A.P., Quaintrell H., Still M.D., Roche P., Charkrabarty D.,
Levine S.E., 1997, MNRAS, 288, 43

\bibitem[]{}
Smale A.P., Kuulkers E., 1999, ApJ, in press, ({\bf astro-ph/9907303})

\bibitem{}
Sood R., Durouchoux P., Campbell-Wilson D., Vilhu O., Wallyn P., 1997,
In : Proc. 2nd INTEGRAL workshop `The Transparent Universe', ESA SP-382,
p.201

\bibitem[]{}
Spencer R.E., Tzioumis A.K., Ball L.R., Newell S.J., Migenes V., 1997,
Vistas Astron., 41, 37

\bibitem[]{}
Stirling A., Spencer R., Garrett M., 1998, New Astronomy Reviews, 42, 657

\bibitem{}
Svensson R., 1998, in Abramowicz M., Bjornsson G., Pringle J.E. (Eds),
{\em Theory of black hole accretion discs}, Cambridge Contemporary
Astrophysics, CUP, p.284

\bibitem{}
Tanaka Y., Lewin W.H.G., 1995,  in Lewin W.H.G., van Paradijs J., van
den Heuvel E.P.J., eds., {\em X-ray binaries}, CUP, p.126

\bibitem{}
Tananbaum H., Gursky H., Kellogg E., Giacconi R., 1972, ApJ, 177, L5


\bibitem{}
van der Klis M., 1995, in Lewin W.H.G., van Paradijs J., van
den Heuvel E.P.J., eds., {\em X-ray binaries}, CUP, p. 252

\bibitem{}
van der Klis M., 1999, In: Proceedings of the Third William Fairbank
Meeting, Rome, in press, ({\bf astro-ph/9812395})

\bibitem{}
van Paradijs J., 1995, in Lewin W.H.G., van Paradijs J., van
den Heuvel E.P.J., eds., {\em X-ray binaries}, CUP, p.536 [vP95]

\bibitem{}
van Paradijs J., White N., 1995, ApJ, 447, L33

\bibitem{} 
White N.E., Nagase F., Parmar A.N., 1995, in Lewin W.H.G.,
van Paradijs J., van den Heuvel E.P.J., eds., {\em X-ray binaries},
CUP, p.1

\bibitem[]{}
Wijnands R., van der Klis M., 1998, Nature, 394, 344

\bibitem[]{}
Zhang S.N., Cui W., Harmon B.A., Paciesas W.S., Remillard R.E., van
Paradijs J., 1997, ApJ, 477, L95

\bibitem{}
Zwarthoed G.A.A., Stewart R., Penninx W., van Paradijs J., van der
Klis M., Roy A.L., Amy S.W., 1993, A\&A, 267, 101

\end{thebibliography}
\end{document}